\documentclass[a4paper,american]{scrartcl}
\usepackage{lmodern}
\usepackage[T1]{fontenc}
\usepackage[utf8]{inputenc}
\usepackage{babel}
\usepackage{amsmath}

\usepackage[authoryear,round]{natbib}
\usepackage[indentfirst=false,font=small]{quoting}
\usepackage{latexsym}
\usepackage[unicode=true,pdfusetitle, bookmarks=false, breaklinks=false,pdfborder={0 0 0},backref=false,colorlinks=false]{hyperref}

\renewcommand{\vec}{\boldsymbol}

\begin{document}

\title{Quantity of Matter or Intrinsic Property: Why Mass Cannot Be Both}

\author{Mario Hubert\thanks{Université de Lausanne, Faculté des lettres, Section de philosophie,
1015 Lausanne, Switzerland. E-mail: \protect\href{mailto:Mario.Hubert@unil.ch}{Mario.Hubert@unil.ch}}}
\maketitle
\begin{abstract}
I analyze the meaning of mass in Newtonian mechanics. First, I explain the notion of primitive ontology, which was originally introduced in the philosophy of quantum mechanics. Then I examine the two common interpretations of mass: mass as a measure of the quantity of matter and mass as a dynamical property. I claim that the former is ill-defined, and the latter is only plausible with respect to a metaphysical interpretation of laws of nature. I explore the following options for the status of laws: Humeanism, primitivism about laws, dispositionalism, and ontic structural realism.

\medskip{}

\noindent \emph{Keywords}: mass; Newtonian mechanics; primitive ontology; Humeanism; primitivism about laws; dispositionalism;
ontic structural realism.
\end{abstract}

\section{Primitive Ontology\label{sec:Particles-as-Primitive-Stuff}}

Any scientific theory must explicitly state what it is about. In particular, every fundamental physical theory must explain the aspect of the world to which its mathematical formalism refers. Albert Einstein reminds us of this truism:
\begin{quote}
Any serious consideration of a physical theory must take into account
the distinction between the objective reality, which is independent
of any theory, and the physical concepts with which the theory operates.
These concepts are intended to correspond with the objective reality,
and by means of these concepts we picture this reality ourselves.
\citeyearpar[p.\ 777]{Einstein:1935gf}
\end{quote}

This seemingly innocent quote contains a strong metaphysical claim and a non-trivial epistemological assertion. On the one hand, Einstein presupposes a world existing in- dependently of any human being. There is an objective reality irrespective of the way we perceive or make judgments about it. On the other hand, we can form physical theories in order to account for the behavior of objects in the world. Physics, in particular, uses mathematics as its central language. And here lies the challenge physics has to meet, since the mathematical entities, like numbers or functions, do not refer to anything in the world unless they are interpreted as doing so. An even greater problem arises when the mathematical entities refer to objects that cannot be directly perceived by our sense organs, for there is always a grain of doubt about their existence.

But physical theories, by means of mathematics, are all we have to explain and predict the behavior of objects, such as electrons, tables, stars, and galaxies. And not all the mathematical entities of a physical theory stand on an equal footing. First, the theory has to postulate basic material entities that are supposed to be the constituents of all the objects around us. Without this requirement, a physical theory is empty.

This requirement for any fundamental physical theory started to get lost in the formation of quantum mechanics. As such, Einstein continued to call attention to it in his philosophical writings. And more than half a century later, the mathematical physicists Dürr, Goldstein, and Zanghì formed the notion of a \emph{primitive ontology} \citeyearpar[original paper from 1992 reprinted in][Chap.\ 2]{Durr:2013fk} to remind us that quantum mechanics has to postulate certain basic objects in order to be a meaningful theory.

By definition, a primitive ontology consists of the fundamental building blocks of matter in three-dimensional space. It cannot simply be inferred from the mathematical formalism of the theory. Instead, it must be postulated as its referent. So all objects, like tables and chairs, are constituted by the elements of the primitive ontology, and the behavior of these elements determines the behavior of the objects.  \citet{Maudlin:2015aa} emphasizes that with the help of a primitive ontology, a physical theory establishes a connection between theory and data. In particular, every measurement-outcome can eventually be explained in terms of a primitive ontology, and the measurement apparatus has no special status with respect to the measured system.

So what do the elements of a primitive ontology look like? This depends on the physical theory we use. In quantum mechanics, for instance, there are three famous options, which actually lead to three different \emph{theories} and not only to three different \emph{interpretations} of the same theory: Bohmian mechanics presupposes a particle ontology; GRWm, a continuous distribution of matter; and GRWf, flashes, that is, a discrete distribution of events in space-time. As in the Bohmian case, the primitive ontology of Newtonian mechanics consists only of particles.\footnote{I consider Newtonian mechanics an action-at-a-distance theory. Interpreting this theory as postulating a gravitational field in addition to the particles would open Pandora's box about the ontological status of fields in general. A detailed treatment of classical fields is
beyond the scope of this paper.} Particles are point-size objects sitting on points of Newton’s absolute space. A point in space can either be occupied by a particle, or it can stay empty. And therefore two or more particles cannot share the same point in space at the same time.

In order to account for the behavior of the primitive ontology, a physical theory has to introduce \emph{dynamical} entities. The predominant dynamical elements of Newtonian mechanics are mass and forces. The standard story is that particles have mass, and in virtue of having mass they exert certain forces between one another. Mass and forces play a different role to particles. While particles constitute all physical objects, mass and forces constrain the motion of the particles.

In this paper, I focus on the ontological role of mass---forces will be treated only in so far as they elucidate the role of mass. There are two standard ways to interpret the ontological status of mass: it can be the measure of the quantity of matter or an intrinsic property of particles. I argue in the next section that the quantity of matter has to be defined in a different way. In Section \ref{sec:dynamical-property}, I explain that the status of properties depends on the metaphysics of laws of nature. There are three predominant positions: Humeanism, primitivism about laws, and dispositionalism.  In each theory mass plays a different ontological role.

\section{Mass and Quantity of Matter}

Newton starts his \emph{Mathematical Principles of Natural Philosophy}
with a definition of the quantity of matter:
\begin{quote}
\textbf{Definition 1}

\noindent \emph{Quantity of matter is a measure of matter that arises
from its density and volume jointly.}

If the density of air is doubled in a space that is also doubled,
there is four times as much air, and there is six times as much if
the space is tripled. The case is the same for snow and powders condensed
by compression or liquefaction, and also for all bodies that are condensed
in various ways by any causes whatsoever. {[}\ldots{}{]} Furthermore,
I mean this quantity whenever I use the term ``body'' or ``mass''
in the following pages. It can always be known from a body's weight,
for---by making very accurate experiments with pendulum---I have found
it to be proportional to the weight, as will be shown below. \citeyearpar[pp.\ 403--404]{Newton:1999aa}
\end{quote}
If mass is defined as density times volume, then the notion of mass has no physical content or explanatory value, since the density itself is defined as mass per volume. Ernst Mach harshly criticizes Newton’s definition on this point:
\begin{quote}
\noindent Definition 1 is, as has already been set forth a pseudo-definition.
The concept of mass is not made clearer by describing mass as the
product of the volume into density as density itself denotes simply
the mass of unit volume. The true definition of mass can be deduced
only from the dynamical relations of bodies. \citeyearpar[p.\ 241]{Mach:1919aa}
\end{quote}
Newton does not give a definition of density; nor is density examined in the scholium following the definitions. It seems that Newton assumes that the reader has a pre-knowledge or an intuition about density such that Definition 1 is more of a rule showing how mass, volume, and density are related rather than a logical definition. As Mach correctly states, a definition of mass in the above sense does not work, which leads to the following two questions:
\begin{enumerate}
\item What does ``quantity of matter'' mean?
\item Is mass connected to the quantity of matter?
\end{enumerate}

Concerning the first question, a primitive ontology of particles allows us to count the particles in a certain volume, and it is natural to take this as the \emph{definition} of quantity of matter without getting into any redundancy. For instance, the quantity of matter of a table consists then of the number of particles, which form the table. 

Since the particles themselves have no internal structure, it does not make sense to assign a quantity of matter to each. At least, it is not meaningful to assign different quantities of matter to particles so that a particle $A$ carries a quantity of matter $a$, and particle $B$ carries a quantity of matter $b$, with  $a\neq b$.

Besides, physics in general and Newtonian mechanics in particular do not need a separate or independent notion of quantity of matter in order to be applied to the world. Statistical mechanics, which relies on counting the number of particles, does not run into conceptual or empirical problems despite lacking an additional notion of the quantity of matter. So it is more precise and parsimonious to define quantity of matter by the number of particles.

Concerning the second question, Newton himself confesses in the last sentence of the quote above that we only have epistemic access to mass when weighing an object, and from the weight we can deduce the quantity of matter. Mach goes a step further, stating that ``[t]he true definition of mass
can be deduced only from the dynamical relations of bodies.'' In this regard, mass is not related to the quantity of matter of an object. Instead, its true and only meaning is dynamical.

\section{Mass as a Dynamical Property\label{sec:dynamical-property}}

The dynamical role of mass is captured in Newton's first and second
law of motion.
\begin{quote}
\textbf{Law 1}

\noindent \emph{Every body perseveres in its state of being at rest
or of moving uniformly straight forward, except insofar as it is compelled
to change its state by forces impressed}.

\medskip{}

\noindent \textbf{Law 2}

\noindent \emph{A change in motion is proportional to the motive force
impressed and takes place along the straight line in which that force
is impressed}. \citeyearpar[p.\ 416]{Newton:1999aa}
\end{quote}
The first law states that the natural motion of a particle is inertial motion, that is, either staying at rest or moving with constant velocity in a straight line. The only thing that can change this motion is the influence of external forces. The second law then shows exactly how the forces act on the particle: first, the stronger the force the greater the acceleration, and, second, the acceleration is parallel to the external force.

Newton’s second law is nowadays mathematically formulated as a differential equation. Consider $N$ particles $P_{1},\ldots,P_{N}$
at positions $\vec{q}_{1}\dots,\vec{q}_{N}$; their trajectories $\vec{q}_{1}(t),\dots,\vec{q}_{N}(t)$
fulfill the differential equation

\begin{equation}
\vec{F}_{i}\left(\vec{q}_{1}(t),\dots,\vec{q}_{N}(t),\dot{\vec{q}}_{1}(t),\dots,\dot{\vec{q}}_{N}(t),t\right)=m_{i}\ddot{\vec{q}}_{i}(t),\label{eq:Newton Second Law}
\end{equation}
where $\vec{F}_{i}$ is the force on the $i$-th particle, $\dot{\vec{q}}_{i}$
its velocity, $\ddot{\vec{q}}_{i}$ its acceleration, and $m_{i}$
its inertial mass. Clearly, the above differential equation makes precise what Newton put into words. And it includes the content of his first law, too: the absence of forces results in inertial motion. So there is actually only one law of motion that generates all classical trajectories of particles, namely, the above differential equation~(\ref{eq:Newton Second Law}).

Still, the law of motion is not complete. We need a precise formulation of the forces involved. On the fundamental level,
one important force is gravitation:

\begin{equation}
\vec{F}_{i}\left(\vec{q}_{1},\dots,\vec{q}_{N}\right)=\sum_{j\neq i}\mathrm{G}\,m_{i}m_{j}\frac{\vec{q}_{j}-\vec{q}_{i}}{\left\Vert \vec{q}_{j}-\vec{q}_{i}\right\Vert ^{3}}\label{eq:gravitational force}
\end{equation}
with the gravitational constant $\mathrm{G}$ and the gravitational
masses $m_{i}$ and $m_{j}$ of the particles $P_{i}$ and $P_{j}$
respectively. The inertial and gravitational masses are a priori physically distinct quantities: the former is a feature of all particles and must be considered in all kinds of interactions; the latter is a specific quantity as part of the law of gravitation  (\ref{eq:gravitational force}).
It is an empirical fact that inertial mass equals gravitational mass, and, therefore, we can treat them as one quantity. Note also that in Newton’s theory there are no massless particles, because equation (\ref{eq:Newton Second Law})
breaks down if we insert $m=0$. So mass is an essential feature of
particles in Newtonian mechanics.

Construed as a dynamical property, there are three ways in which physics describes
mass:
\begin{enumerate}
\item mass is an intrinsic property of particles;
\item mass is just a parameter of the laws of motion;
\item mass is a coupling constant.
\end{enumerate}

I claim that these three interpretations can only be made precise with respect to some metaphysical framework. In what follows, we discuss mass in the framework of Humeanism, primitivism about laws, and dispositionalism. Finally I interpret mass within the theory of ontic structural realism, which I regard as an instance of dispositionalism.

\subsection*{Humean Supervenience}

Humean supervenience, the modern form of Hume’s metaphysics, was first posited by David Lewis:
\begin{quote}
It is the doctrine that all there is to the world is a vast mosaic
of local matters of fact, just one little thing and then another.
{[}\ldots{}{]} We have geometry: a system of external relations of
spatiotemporal distance between points. Maybe points of spacetime
itself, maybe point-sized bits of matter or aether fields, maybe both.
And at those points we have local qualities: perfectly natural intrinsic
properties which need nothing bigger than a point at which to be instantiated.
For short: we have an arrangement of qualities. And that is all. All
else supervenes on that. \citeyearpar[pp.\ ix--x]{Lewis:1986aa}
\end{quote}
The ontology of Humean supervenience is characterized by the contingent distribution of local matters of particular facts: the Humean mosaic. There is a net of spatiotemporal points that are connected only by external metrical relations, and, at those points, certain qualities can be instantiated by at least one of three entities that Lewis regards as fundamental: space-time itself, particles, or values of fields. Given some initial distribution of those qualities there is nothing in the ontology that constrains its further development.

Obviously, our world contains regularities, but on the Humean view this is just a contingent fact. In order to avoid giving an enormously long list of particular facts describing these regularities, \citet[p.\ 478]{Lewis:1994aa}
introduces his best system account of laws of nature. According to his proposal, the laws of nature are theorems of the best deductive system, which combines or balances simplicity and strength in describing the temporal development of local matters of particular facts throughout space and time. A long list of these facts would be highly informative but very complex, whereas a single law of nature would be very simple but probably not contain enough information. So the best system comprises a certain finite number of laws of nature as theorems, which offer the perfect compromise.

In Lewis’s Humeanism, mass can be part of the ontology of the mosaic: in which case it is a ``natural intrinsic property'' instantiated at points of space-time. This move, however, poses a serious metaphysical problem: mass becomes a categorical property that is defined as independent of the causal role it plays in the world. Hence, mass has a primitive identity or quiddity, which allows it to play a different causal role in another possible world. For example, in another possible world mass could play the role of charge. This would be the very same property that we call mass, but it would act like charge does in our world.

This seems absurd and leads to the problem---called humility---of our not having epistemic access to the true identity of mass, because all we can know are the causal consequences of properties. So, given two worlds that coincide in the temporal development of all their particles, it would, first, be metaphysically possible for these worlds to be different with respect to the quiddity of their categorical properties, and, second, it would be impossible for us to know which world we inhabited.  Lewis bites the bullet and accepts this metaphysical burden in favor of a sparse ontology with no modal connections.

Ned \citet[Sec.\ 5.2]{Hall:2009aa} proposes a different strategy for conceptualizing the status of mass. He interprets the Humean mosaic as consisting solely of point-sized particles standing in certain spatiotemporal relations \citep[see also][]{Loewer:1996aa,Esfeld:2014ab}. 
The particles do not have intrinsic properties, let alone categorical ones, and all non-modal facts about the world are just the positions of these particles. Mass enters the scene as part of
the best system describing the temporal development of the particles, as part of the fundamental laws of nature in the Humean sense, and as part of the differential equations that describe the trajectories of particles. In a description of the history of the world that balances simplicity and informativeness, mass functions as a parameter in this best system. 

For Newtonian mechanics, restricted to gravitational interaction, the best system may be interpreted as consisting of the equations~(\ref{eq:Newton Second Law})
and~(\ref{eq:gravitational force}).
As a way of speaking or as a convenient metaphor, we can ascribe these parameters to the particles themselves such that every particle $P_{i}$ is characterized by a magnitude $m_{i}$. But this interpretation of mass does not change or add anything to the ontology. The distribution of propertyless particles is the entire ontology; everything else, including mass, supervenes on this mosaic. As such, there are no categorical properties in the ontology, and the problem of quiddity or humility does not arise in this version of Humean supervenience.

One general critique attacks Humean supervenience on a point that Humeans regard as one of its greatest virtues: the sparse ontology that lacks modal connections. It is unsatisfactory that there are no facts about \emph{why} we see regularities in our world. 
Even the laws of nature as part of the best system cannot explain \emph{why} particles follow a Newtonian trajectory. For particles just move as they do, in a contingent way. All a Humean can do is give a good description or summary of the regularities, and if the regularities change she has to change her description too. Consequently, we have to include modal connections in the ontology.

\subsection*{Primitivism about Laws}

Primitivism about laws regards the existence of the laws of nature as a primitive fact, where the laws themselves govern the behavior of the primitive ontology. One famous adherent of this position is Tim Maudlin:
\begin{quote}
To the ontological question of what makes a regularity into a law
of nature I answer that lawhood is a primitive status. Nothing further,
neither relations among universals nor role in a theory, promotes
a regularity into a law. {[}\ldots{}{]} My analysis of laws is no
analysis at all. Rather I suggest we accept laws as fundamental entities
in our ontology. Or, speaking at the conceptual level, the notion
of a law cannot be reduced to other more primitive notions. The only
hope of justifying this approach is to show that having accepted laws
as building blocks we can explain how our beliefs about laws determine
our beliefs in other domains. Such results come in profusion. \citeyearpar[pp.\ 17--18]{Maudlin:2007aa}
\end{quote}
As stated by Maudlin, the entire ontology is made up of the primitive ontology plus the laws of nature. It is a primitive fact that there are laws of nature, and that particles move according to these laws. The task of physics, then, is to discover these laws. For instance, it is a primitive fact that equations (\ref{eq:Newton Second Law})
and (\ref{eq:gravitational force}) hold in a Newtonian universe, and here we come to an answer regarding \emph{why} a particle follows a Newtonian trajectory: because there are such laws.

What is the role of mass in this framework, then? It is just a parameter of the Newtonian laws of motion referring to nothing at all in the primitive ontology. Mass is not a parameter that results from the best description, as in the Humean case; rather it is an essential parameter of the laws of nature leading to correct trajectories.

The notion of a parameter is slightly inappropriate here, because it invites us to think about mass as being adjusted or altered under certain circumstances. But the only circumstance available to us is the universe as a whole. There is a primitive ontology consisting of N particles and the laws of nature. And it happens to be the case that the law is formulated such that there are $N$ constants $m_{1},\ldots,m_{N}$. So it seems more appropriate to interpret $m_{1},\ldots,m_{N}$ as \emph{constants of nature} on a par with the gravitational constant $\mathrm{G}$ or Planck's constant $\hbar$. Recognizing masses as constants of nature is clearer and more in the spirit of primitivism than dubbing them parameters. It is likely that the idea of mass as a parameter came from the application of Newton’s laws to real life cases, where it had to be adjusted to describe the physical bodies of a given subsystem.

Similarly to Hall, one can pretend that the parameter mass is ``located'' at the particle’s position and speak as if it were an intrinsic property of particles. In this sense mass still has a purely nomological role, but this way of speaking may aid our intuition.

Primitivism about laws retrieves modal connections as part of ontology in the form of laws. Maudlin does not state how laws are connected to the primitive ontology. There seems to be an intuition that laws ``govern'' or ``direct'' the behavior of particles, but these phrases are purely metaphorical \citep[p.\ 119]{Loewer:1996aa}. In the above quote, there is no attempt to explain these metaphors: ``lawhood is a primitive status.'' Nevertheless, one can ask, ``How can a law as an abstract entity govern anything in the world? How can particles or any material body ‘obey’ these laws?'' Primitivism about laws just answers, ``It is a primitive fact.'' Nonetheless, one position that tries to answer these questions by introducing an underlying mechanism is dispositionalism.

\subsection*{Dispositionalism}

This strategy tries to recover modal connections by introducing dynamical properties into physical systems, which are called \emph{dispositions} or \emph{powers} \citep[for instance,][]{Bird:2007aa}. Accordingly, a physical system behaves the way it does because it has a certain property or disposition to do so. This idea can be applied to the primitive ontology of Newtonian mechanics. Mass is then an intrinsic property of particles.  It is intrinsic in the sense that the mass of one particle does not depend on the masses of other particles.

Mass, interpreted as a disposition, does not give an intrinsic identity to particles. The identity of particles stems from their location in space. Since Newtonian mechanics relies on an absolute background space, where every point in space is by definition distinguished from any other point in space, it is sufficient to ground the identity of particles on their position in absolute space.\footnote{As argued in \citet{Esfeld:2014ac}, this can be already done in a relational space.} 
The role of mass is solely a dynamical one; that is to say, it constrains the motion of particles.

Moreover, it is essential for mass to have the same causal-nomological role in all possible worlds; it is not a categorical property, and consequently it does not bear the problems of either quiddity or humility. 
This causal-nomological role is expressed by Newton’s laws (\ref{eq:Newton Second Law}) and (\ref{eq:gravitational force}). 
In other words, Newton’s laws are grounded in the ontology by the intrinsic masses of particles. Our epistemic access to mass as a disposition is possible through observation of what it does in the world, that is, its causal-nomological role; and the laws of nature  (\ref{eq:Newton Second Law}) and (\ref{eq:gravitational force}) are a concise expression of its effects.

A crucial feature of dispositions is their need for certain triggering conditions in order to be manifested in the world. 
Zooming into Newton’s second law (\ref{eq:Newton Second Law}) we can see the following: the manifestation of the mass $m_{j}$ of particle $P_{j}$ at time $t$ is its acceleration $\vec{a}_{j}(t)$ given the positions and velocities of all the particles (including $P_{j}$) at time $t$. So the positions and velocities of \emph{all} particles are triggering conditions for the manifestation of mass. In the case of gravitation, the mass $m_{j}$ cancels out on both sides of  (\ref{eq:Newton Second Law}), and we deduce that the acceleration $\vec{a}_{j}(t)$ does not mathematically depend on $m_{j}$. Yet, $\vec{a}_{j}(t)$ is the manifestation of the mass of the $j$-th particle, though it is independent of the precise value of $m_{j}$.

\subsection*{Ontic Structural Realism}

The interpretation of mass as a coupling constant does not seem to fit either of the metaphysical schemes discussed above. What is the ontological status of mass distinct from its being an intrinsic property of particles or a constant in the laws of motion? 
Mass, interpreted as a \emph{coupling} constant, emphasizes the dynamical relations between particles. Particles move as they do because they stand in certain relations described by Newton’s laws (\ref{eq:Newton Second Law}) and (\ref{eq:gravitational force}), and the role of mass is then to quantify these relations.

A metaphysical approach that supports this view is ontic structural realism (OSR). According to the original idea of OSR, the world consists purely of structures, all 
the way down to the fundamental level \citep{Ladyman:2007aa,French:2014aa}. 
If there happen to be physical objects in the ontology, they are interpreted as nodes of structures. And this is the weak point of OSR, because the existence of structures without objects to instantiate them is implausible. \citet{Esfeld:2009aa} therefore suggests that OSR requires objects as the relata of structures, and he interprets the structures as being modal. That is to say, they constrain the temporal development of the objects instantiating them.

Esfeld’s proposal qualifies OSR as an instance of dispositionalism. The only difference lies in the nature of the dynamical entities. Intrinsic properties are no longer responsible for the dynamical constraints; this task is fulfilled solely by relations between the elements of the primitive ontology.

It is now straightforward to apply this idea to Newtonian mechanics. Particles are the objects that stand in certain spatiotemporal relations resulting from their positions in absolute space, and in addition to these spatiotemporal relations they stand in certain dynamical relations. 
The latter relations are the modal structure. In the case of gravitation, this structure functions according to (\ref{eq:gravitational force}), and the manifestation of this structure is the acceleration of particles according to (\ref{eq:Newton Second Law}). Note that the spatial relations between the particles are not modal, because these relations alone have no causal-nomological role in the dynamical behavior of particles.

So mass cannot be interpreted as an intrinsic property of particles in this framework; rather, it is  a  parameter  that  specifies  the  gravitational  structure  regarded  as an additive bipartite particle–particle relation according to 
(\ref{eq:gravitational force}), and, in this sense, the particles are coupled. In other words, the motion of one particle changes the motion of other particles in the universe because taken together particles instantiate a dynamical structure. 
A crucial feature of this dynamical structure is that it is reducible to or separable into direct relations between two particles; this reduction fails in the quantum case, which requires a non-separable holistic structure as proposed by
\citet{Esfeld:2012mz}. In sum, the notion of a coupling constant points to two aspects of mass: on the one hand, mass is a constant in the laws of motion, and, on the other hand, this notion anticipates dynamical relations between particles.

\section{Conclusion}

The aims of this paper were twofold. First, I showed that the notion of a primitive ontology can be fruitfully used in classical mechanics. Second, I argued that the status of mass depends on the metaphysics of the laws of nature. It subsequently became clear that mass has to be interpreted as a dynamical entity introduced by Newton’s laws of motion. The metaphysical theories that I discussed allow mass to be construed in three different ways: it may be regarded as a parameter, as an intrinsic property, or as a coupling constant. I tried to remain neutral with respect to the ``best'' interpretation. A thorough evaluation of the different positions remains to be undertaken.

\section*{Acknowledgments}

I wish to thank Michael Esfeld, Dustin Lazarovici, and an anonymous referee for many 
helpful comments on previous drafts of this paper. 
This work was supported by the Swiss National Science Foundation, grants no.\ PDFMP1\_132389.

\bibliographystyle{abbrvnat}
\bibliography{references}

\end{document}